\documentclass{SCGE}
\usepackage{rotating}

\begin{document}

%%%%%%аæʽҪ¼ÓÉÏÕâ×é
\begin{picture}(0,0){\rm
\put(0,-20){\makebox[160truemm][l]{\bf {\sanhao\raisebox{2pt}{.}}
Article  {\sanhao\raisebox{1.5pt}{.}}}}}
\put(0,-34){\jiuwuhao {\textcolor[rgb]{0.5,0.5,0.5}{\sf %Progress of Projects Supported by NSFC
}}}%%(11ÔÂ×¢ÊÍ£ºµ÷\textcolor[rgb]{x,x,x}ÖеÄÊý×ÖxÔ½´óÔ½»Ò)
\end{picture}

\def\bm{\boldsymbol}

\def\dl{\displaystyle}
\def\du{\end{document}}
\def\d{{\rm d}}
\def\e{{\rm e}}
\def\i{{\rm i}}

\def\pi{{\uppi}}

% The author doesn't need fill in it.
\Year{2015} %
\Month{April} %
\Vol{58} %  ¾íºÅ
\No{1} %  ÆÚºÅ
\BeginPage{1} % ÆðÒ³Âë
\AuthorMark{{\rm Yi W.}, et al.}  %(11ÔÂ×¢ÊÍ£ºÒ³Ã¼ÉϵÄ×÷Õß)
\AuthorMarkCite{{\rm Yi W., et al}.~} %(11ÔÂ×¢ÊÍ£ºcitationÖеÄ×÷Õß)
\DOI{10.1007/s11433-014-5625-8} % The author doesn't need fill in it.
\ArtNo{012002}

% \title[short text for running head]{full title}{comments for title}
\title[Discovery of two BALQSOs at $z \sim 4.75$]{Discovery of two broad absorption line quasars at redshift about 4.75 using the Lijiang 2.4m telescope}

\author[1,2,3]{Yi Weimin}{}
\author[4,5*]{Wu Xue-Bing}{}
\author[4]{Wang Feige}{}
\author[4]{Yang Jinyi}{}
\author[4]{Yang Qian}{}
\author[1,3]{Bai Jinming}{}

\address[{\rm1}]{Yunnan Observatories, Chinese Academy of Sciences, Kunming 650011,China}
\address[{\rm2}]{University of the Chinese Academy of Sciences, Beijing 100049, China}
\address[{\rm3}]{Key Laboratory for the Structure and Evolution of Celestial Objects, Chinese Academy of Sciences, Kunming 650011,China}
\address[{\rm4}]{Department of Astronomy, School of Physics, Peking University, Beijing 100871, China}
\address[{\rm5}]{Kavli Institute for Astronomy and Astrophysics, Peking University, Beijing 100871, China}

\maketitle \vspace{-3.5mm}{\footnotesize\begin{center} Received xx, xxxx; accepted xx, xxxx; published online xx, xxxx
\end{center}}\vspace*{-5mm}

%     Abstract is required.
\begin{center}
\rule{16.5cm}{0.4pt}
\parbox{16.5cm}
{\begin{abstract} 
The ultraviolet broad absorption lines have been seen in the spectra of quasars at high redshift, and are generally considered to 
be caused by outflows with velocities from thousands kilometers per second to one tenth of the speed of light. They provide crucial 
implications for the cosmological structures and physical evolutions related to the feedback of active galactic nuclei (AGN). Recently, 
through a dedicated program of optical spectroscopic identifications of selected quasar candidates at redshift 5 by using the Lijiang 2.4m 
telescope, we discovered two luminous broad absorption line quasars (BALQSOs) at redshift about 4.75. One of them may even have the 
potentially highest absorption Balnicity Index (BI) ever found to date, which is remarkably characterized by its deep, broad absorption 
lines and sub-relativistic outflows. Further physical properties, including the metal abundances, variabilities, evolutions of the 
supermassive black holes (SMBH) and accretion disks associated with the feedback process, can be investigated with multi-wavelength 
follow-up observations in the future. 

\end{abstract}}
\end{center}\vspace*{-0.6cm}

\begin{center}
\parbox{16.5cm}
{\bf\jiuhao high-redshift, active galactic nuclear (AGN),  broad absorption line quasar (BALQSO), outflow.}%¹Ø¼ü´Ê
\end{center}

\begin{center}
{}%·ÖÀàºÅ
\CITA    %%(11ÔÂ×¢ÊÍ£ºCitationÄÚÈÝ×Ô¶¯Éú³É)
%\Cit{~~~???, et al. ???. Sci China-Phys Mech Astron, 2014, 57: 1--6, doi:}%%(11ÔÂ×¢ÊÍ£ºCitationÄÚÈÝÐèÊÖ¶¯Ìîд)
\end{center}

\textwidth=178truemm \textheight=236truemm%%%%%%аæʽҪ¼ÓÉÏ

%%%%%%%%%%%%%%%%%%%%%%%%%%%%%%%%%%%%%%%%%%%%%%%%%%%%%%%%%%%%
\wuhao\vspace*{1.5mm}

\begin{multicols}{2}

%%%%%%%%%%%%%%%%%%%%%%%%%%%%%%%%%%%%%%%%%%%%%%%%%%%%%%%%%%%%
%% Text of article.
%%%%%%%%%%%%%%%%%%%%%%%%%%%%%%%%%%%%%%%%%%%%%%%%%%%%%%%%%%%%
%    Section headings
\renewcommand{\baselinestretch}{1.08} \baselineskip 12.2pt\parindent=10.8pt

\renewcommand{\thefootnote}

\section{Introduction}
Broad absorption line quasars (BALQSOs) are quasars whose UV/optical spectra show absorption troughs blueward from the corresponding 
emission lines with Balnicity Index (BI, \cite{Weymann91}) greater than zero, which in general, occurs in the high-ionization 
transitions of C IV, Si IV, N V, and O IV \cite{Stocke92}. 
Previous spectral studies have indicated that this sub-category comprises 10\% - 20\% of the whole quasar population at low and 
intermediate redshifts (e.g., \cite{Reichard03b,Hewett03,Knigge08,Scaringi09}), characterized by deep and broad absorption 
features associated with UV resonant lines. These BALs are at least 2000 km s$^{-1}$ wide and can be accelerated up to velocities 
of $\sim$0.1c. 
Under this context, high-redshift BALQSOs may provide an unique tool to investigate the circumnuclear medium and corresponding AGN feedbacks, 
evolutionary phases of metal enrichment in the early Universe. 

BALs with deep troughs are found only in the spectra of radio-quiet quasars, never in the spectra of strong radio sources yet \cite{Stocke92}.
The formation of broader and higher velocity absorption troughs seems more likely to be intrinsic, but the definite reason is more 
complicated to address. It is generally considered that only the central engine of quasars can feasibly accelerate matters to such high 
velocities observed in BALQSOs. The activity of BALQSOs may be powered by super/sub-Eddington radiation to some extent under the presumption 
that the huge amount of outflows are expelled and accelerated by the central radiation pressure with appearances of deep, broad absorption 
troughs, and sub-relativistic outflows on a long cosmological timescale. The study of broad absorption lines commonly observed in the 
rest-frame ultraviolet (UV) spectra of optically selected QSOs significantly advances our understanding of the structure and 
emission/absorption physics of AGNs.

So far, there are about 5,000 quasars classified as BALQSOs \cite{Trump06,Gibson09}, among which only $\sim$ 50 BALQSOs have been 
confirmed with redshift larger than 4.5, and the high-redshift BALQSOs with broad, deep absorption troughs and sub-relativistic 
detached velocities are very rare to be seen.
In this paper, we report the discovery of two high-redshift BAL quasars at z $\sim$ 4.75, in which both have sub-relativistic 
high-ionization outflows and one may have the highest BI known to date, indicating some extreme activities or unclear physics 
behind them. The paper is organized as follows. In Section 2, we describe the observations from the Lijiang 2.4m telescope (LJT) 
and the related data-reduction process. Through the spectra collected, we discuss the properties of the two BALQSOs in Section 3, 
including the redshift uncertainties, the BI indices and the SEDs. The results are summarized in Section 4. Throughout this paper, 
we adopt a $\Lambda$-dominated flat cosmology with $H_0$ = 70 km s$^{-1}\cdot$Mpc$^{-1}$,  $\Omega_M$ = 0.3 and  $\Omega_{\Lambda}$ = 0.7.

\noindent\rule{2.5cm}{0.4pt}\\[0.1mm]{\qihao *Corresponding author (email:
wuxb@pku.edu.cn)}%ÊÖ

\end{multicols}

\section{Observations and Data Reduction}
\begin{multicols}{2}
\renewcommand{\baselinestretch}{1.08} \baselineskip 12.2pt\parindent=10.8pt
\begin{figure}[H]
 \includegraphics[width=9.0cm, height=6cm, angle=0]{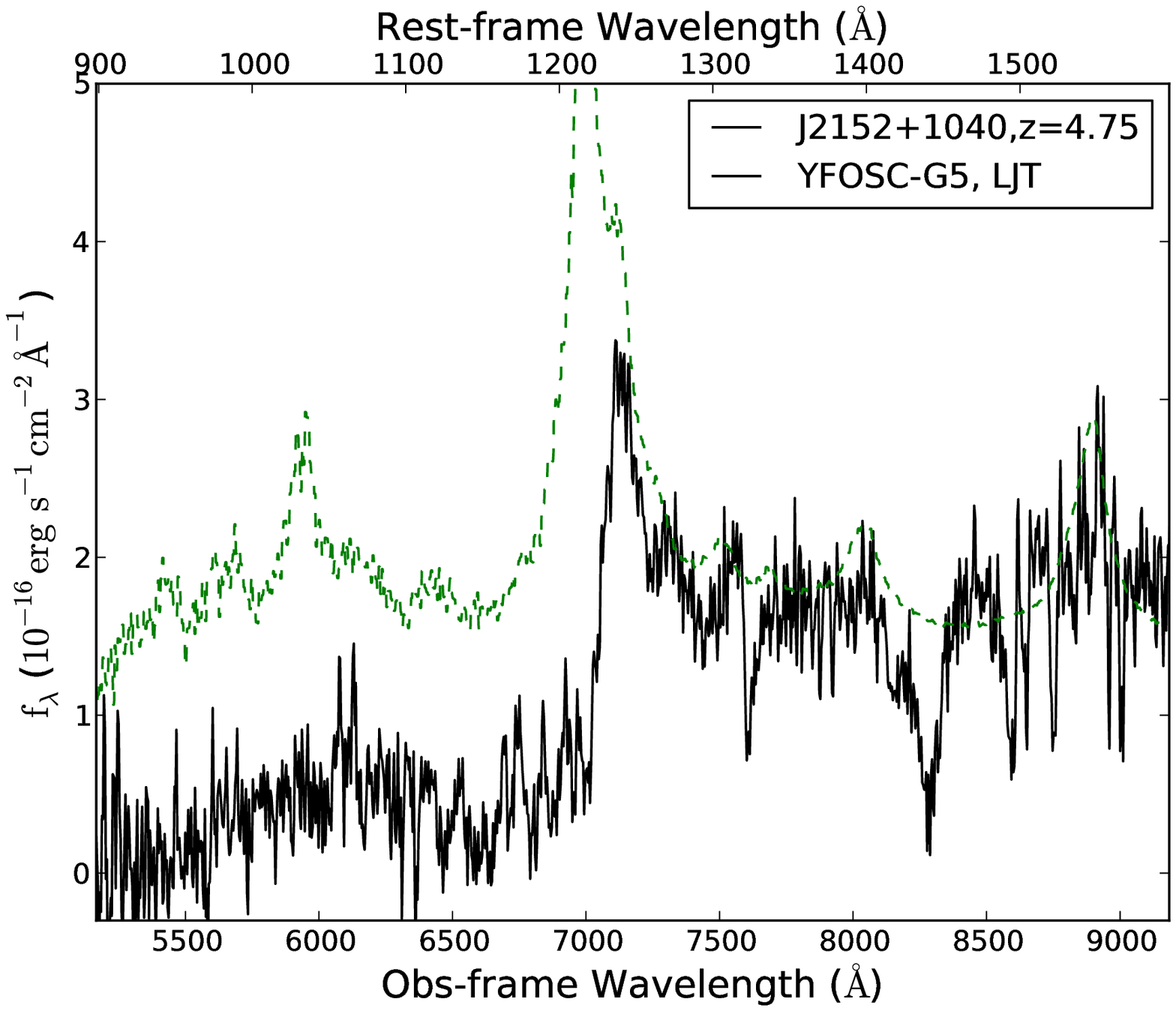}
  \includegraphics[width=9.0cm, height=6cm, angle=0]{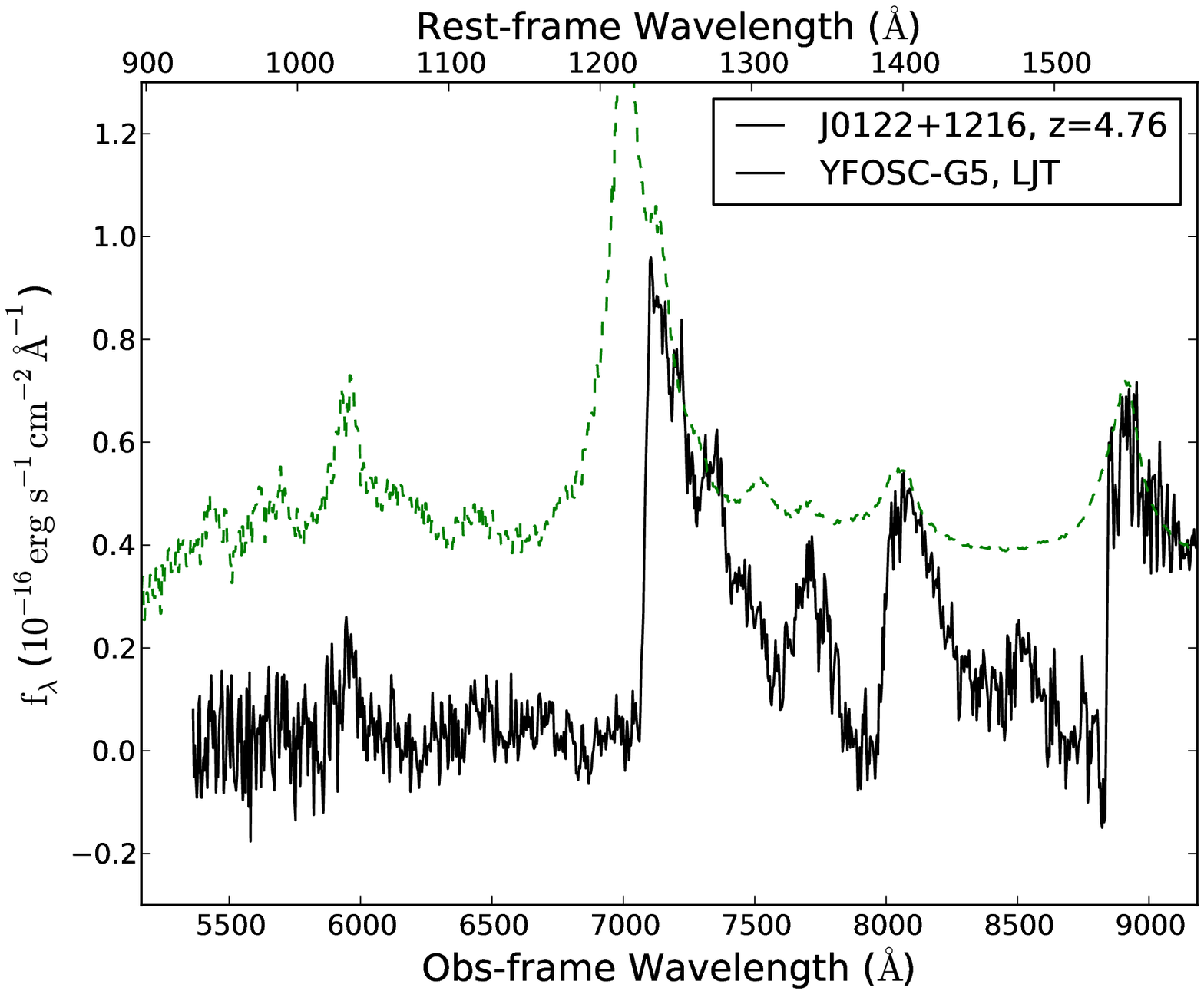}
 \caption{The optical spectra of the two BALQSOs (solid black). The top panel presents the spectrum of J2152+1040 obtained by the 2.4m 
 telescope with YFOSC spectrograph on Oct. 1, 2014. The bottom panel presents the spectrum of J0122+1216 obtained with the same instrument 
 on Oct. 24, 2014. The scaled composite spectrum (dashed green) of SDSS QSOs in each panel is presented as a comparison to show the absorption features
 in the observed spectrum.}
  \label{fig:simple}
\end{figure}
In the end of 2013, we started a quasar-identifying campaign aimed at finding luminous high redshift quasars according to the optical-IR selection 
criteria based on SDSS and WISE photometric data \cite{wxb12}. The two BAL quasars, J215216.09+104052.44 and J012247.34+121624.00 (hereafter 
J2152+1040 and J0122+1216), were discovered during the observational runs conducted with LJT, and the spectroscopic observations were 
carried out on October 1 and October 24, 2014, respectively. Among the various imaging and spectroscopic observing modes, LJT equipped 
with the Yunnan Fainter Object Spectrograph and Camera (YFOSC, \cite{zhang14}), offers a rapid switch, high sensitivity, low resolution 
observing mode, and we use a redward sensitive grism 5 as the dispersing element for the dedicated observations.

These two quasars were found to be BALQSOs with striking broad absorption lines (BALs) at first glance after the spectral extraction. 
Wavelength calibration was performed by using Neon and Helium lamps. The telluric absorption feature at $\sim$7600\AA~ is obvious in 
all of the spectra. Using the spectrum of a spectrophotometric standard star, absolute flux calibration was obtained under a photometric 
night observed at similar airmass. Considering some technical details of LJT and the effects of the unstable seeing, differential atmospheric 
refraction will cause a target's centroid to be extended a little in the vertical direction of the sky, so the dewar (dispersion components 
and CCD attached) was rotated at 90 degrees to make sure our targets' photon energy being fallen into the slit as much as possible.

~\\

During the observations of J2152+1040 and J0122+1216, the 1.8'' slit and grism 5 with a dispersion of 3.6 \AA~ pixel$^{-1}$ were used to 
take their spectra from YFOSC, providing a typical resolution of $R = \lambda/\Delta\lambda \approx$ 550 in the spectral range from 
5500 to 9200 \AA~ \cite{yi14}. The blue and red magnitudes released from the SDSS DR10 are $r = 20.17$ and $i = 18.5$ for J2152+1040,  
$r = 22.35$ and $i = 19.44$ for J0122+1216, respectively. We also note that both quasars are radio quiet for their radio flux below 
the threshold of the FIRST radio survey \cite{Becker95}. Details of photometric magnitudes of the two BALQSOs can be obtained from 
the literature, including the SDSS, 2MASS, {\it WISE} \cite{York00,Wright10,Skrutskie06}, which are listed in Table 1. 

\begin{table*}
\centering
 \caption{Photometric data from multiple surveys and derived observational results}
 \begin{tabular}{lcccr}
  \hline\noalign{\smallskip}
 Name & J0122+1216 & J2152+1040 \\
 SDSS-u & 23.43$\pm$0.59 & 22.97$\pm$0.27 \\
 SDSS-g &  24.29$\pm$0.37 & 24.28$\pm$0.32 \\
 SDSS-r  &  22.35$\pm$0.14  & 20.17$\pm$0.03 \\
 SDSS-i & 19.44$\pm$0.03 &  18.51$\pm$0.02 \\
 SDSS-z & 19.32$\pm$0.06 & 18.32$\pm$0.03 \\
 J &  - & 16.86$\pm$0.14 \\
 H & - & 16.29$\pm$0.21 \\
 K & - & 15.13$\pm$0.00  \\
 W1 & 15.58$\pm$0.05 & 14.64$\pm$0.04 \\
 W2 & 15.03$\pm$0.09 & 14.00$\pm$0.05 \\
 W3 & 11.50$\pm$0.16 & 10.73$\pm$0.08 \\
 W4 & 8.62$\pm$0.00 & 7.94$\pm$0.17 \\
  $\rm A_V^a$ & 0.09 & 0.19  \\
  $\it{BI}$(CIV)$^b$ & 15320 & 5100 \\
  $z^c$ & 4.76 &  4.75 \\
L$_{1350}^d($erg s$^{-1}$) & 9.1$\times$10$^{46}$ & 1.3$\times$10$^{47}$ \\
  \noalign{\smallskip}\hline
\end{tabular}
\\
\tiny{Note: a: Galactic dust reddening and extinction \cite{Schlafly11}.\\
 b: Balnicity index defined by the revised calculation (1) in section 3.2.\\
 c: Redshift value ~~~~  d: The monochromatic luminosity at 1350\AA}
\end{table*}

~\\

\section{Discussion}
\subsection{Redshift Uncertainties}
After extracting the spectra from the raw data, redshifts of the two quasars were initially identified by the strongest emission line peaks 
in the spectra, which yielded redshifts of 4.84 and 4.83 for J2152+1040 and J0122+1216 at first. 
While it has been found that there appears large discrepancies on redshifts when we check all the emission lines in the spectra, which leads 
to the re-identifications of their redshifts using the SDSS non-BALs composite template (the green dashed line) as shown in Figure 1. 
Among the two BALQSOs, mainly due to their strong Ly$\alpha$ forest absorption effects and highly luminous continua, it is relatively easy 
to find the unabsorbed emission peaks and hence the redshifts, even though their true values cannot be accurately established.
Furthermore, considering it is difficult to ascertain an accurate position of the N V peak, we believe the complexity of the blended 
Ly$\alpha$+N V in the two BALs makes it risky to deconvolve the two lines.
The average SDSS non-BALs quasar spectrum was carefully shifted on each observational spectrum until we found the best consistency of these 
emission lines' position. Then, the most likely redshifts of the two BAL quasars were corrected to be 4.75 and 4.76 compared to the previous 
ones mentioned above. While the redshift of J2152+1040 has larger uncertainty since the peaks of Ly$\beta$ and C IV are not in good 
agreement with each other, so we took another higher resolution spectrum ( $\lambda/\Delta\lambda \approx$ 700 at 7000\AA) 
by the LJT to double check this discrepancy, but found no difference except for their intensity level. In addition, the relatively narrow absorption lines sparsely located at two sides of the 
C IV peak may be related to the intervening clouds associated with the intergalactic medium (IGM), which could be further confirmed by 
the near-IR spectroscopy in the future.

~\\

It seems that the two BALQSOs have similar absorption features in the position of the Ly$\alpha$ emission line, suggesting the two quasars 
might experience a special phase or a sudden coincidence during the coevolution of BAL outflows and their powering engines. Although the 
possible relationship of the BAL gas with the BLR is still unclear, we could obtain useful clues based on the studies of some disappearing 
emission lines. One case is the strong NV 1240 absorption in BAL QSOs, which often nearly obliterates the Ly$\alpha$ emission line hence 
adds more uncertainties of their redshifts. The other case is that dust extinction may be an important factor affecting the escape of 
Ly$\alpha$ photons, while at low extinctions, other factors such as neutral hydrogen covering factor and gas kinematics could also be 
effective at inhibiting the escape of Ly$\alpha$ photons.
These obvious absorption troughs, in turn, argues that the BAL gas lies outside the BLR and covers nearly all of the line-emitting 
region as seen along the line of sight. In order to highlight the absorption regions, the non-BALQSO composite spectrum is also used 
as a comparison in Figure 1.

\subsection{The C IV Balnicity Indices}

The first large sample of BALQSOs was analyzed by Weymann \cite{Weymann91}, who defined BALQSOs as quasars exhibiting C IV absorption 
troughs broader than 2000 km s$^{-1}$. 
Determining whether a quasar is a BALQSO, actually may be more complicated since 
BI indicates not only the presence of one or more broad absorption troughs but also the amount of absorption.
In particular, how does one define the true continuum level mixed with emission lines and the systematic redshift 
when there is significant absorption? 
Moreover, the obvious drawback of the ideal BI is that it requires exact knowledge of the quasar's systemic velocity and continuum 
level. However, absorption troughs should evidently arise in BAL outflows (of C IV, Si IV or other species) as opposed to being 
intervening absorbers or intrinsic systems unrelated to the BAL outflows. Among the two quasars, there is no difficulty to classify 
them as BALQSO according to their extremely broad and deep absorption troughs.
We retain the traditional BALQSO definition as both quasars with the {\it balnicity index} (BI) $>$ 0 shown as follows: 
\begin{equation}
   {\rm BI} = \int^{-25000}_{-3000}[1-\frac{f(v)}{0.9}]C{\rm d}v
\end{equation}
$f(v)$ is the normalized flux distribution, $C  = 1$ at trough velocities more than 2000 km $\rm s^{-1}$ from the start of a continuous trough 
with flux density less than 90\% of the continuum, and $C = 0$ elsewhere. The 25000 km s$^{-1}$ blue limit is set to avoid ambiguities 
associated with the Si IV emission and absorption (here we actually set the upper limit up to 26000 km s$^{-1}$ considering all of the 
C IV troughs extending to the right edge of the Si IV line), and 3000 km s$^{-1}$ red limit is chosen to avoid contamination from 
absorption that might not be caused by outflows.
\begin{figure}[H]
\centering
  \includegraphics[width=9cm, height=6.5cm, angle=0]{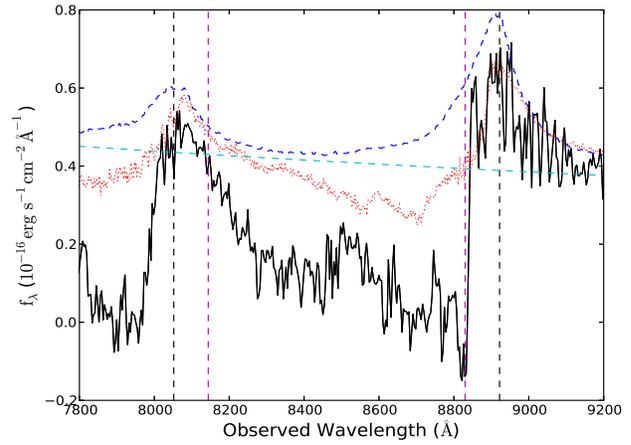}
  \caption{A demonstration of the calculation of C IV BI index for J0122+1216, in which the blue dashed and red dotted lines are scaled to find a better 
  continuum level and represent non-BAL and BAL  composite spectra, respectively. The cyan dashed line shows the well-consisted continua fitting with a 
  power law slope of -1.1. The velocity scale is based on the systematic redshift $z=4.76$. The vertical magenta dashed lines represent the detached 
  velocities from 3000 to 26000 km s$^{-1}$, and the vertical black dashed lines give the observed wavelengths of the C IV and Si IV lines.}
  \label{Fig.1}
\end{figure}

Low-redshift BALQSOs would be better for the quantitative analysis because their BI values are easy to normalize while the continuum fitting of 
high-redshift spectra can be problematic due to significant absorptions \cite{Trump06}. In order to avoid this problem, the best-fit template is 
manually adjusted to prevent overestimation or underestimation of the BI, especially for the quasar owning unusual emission/absorption line profiles. 
In addition, since the Fe emission lines make negligible contributions for the Si IV and C IV lines in the rest-frame wavelengths, we took a 
power-lower function to roughly fit the unabsorbed continua as shown in Figure 2, and the continuum level is set according to the seemingly 
unabsorbed parts of the whole continuum. Therefore, it will give the lower limits of their BI values.
Moreover, the continuum slope and the BI of the C IV that is similar to the equivalent width expressed in km s$^{-1}$, are determined by the 
improved method \cite{Reichard03a}. 

As a result, the calculated BI of J2152+1040 should be close to the true value since its continuum is not absorbed much. For J0122+1216, which 
is conspicuous for its extremely broad and deep absorption troughs, we roughly obtained its BI according to the unabsorbed continua in the scaled 
template. If the continua level of the composite template is higher than the seemingly true continua shown in the observational spectra, these 
scaled continua will be abandoned in the calculation. Thus, the estimated BI indices of the two BALQSOs are close to their lower limits, which 
are tabulated in table 1. For example, after different fittings of the C IV trough for J0122+1216, the results give a lower BI limit of 15320 
km s$^{-1}$, which is possible among one of the highest BI BALQSOs known to date. 
Furthermore, Balnicity indices higher than 8000~km s$^{-1}$
are only found among low ionization BALQSOs (LoBAL). As a comparison, the absorption profile of J0122+1216
is well resolved, in which it not only owns the deepest and broadest absorption troughs, but also displays a clear peak in the middle of the 
C IV double troughs.
This phenomenon, again, is a typical feature commonly observed in LoBALs, so it would be particularly helpful to 
further confirm their nature through near-IR spectroscopic observations in the future.

\subsection{Luminosity Properties Based on SEDs}
Since radio surveys have demonstrated that the decreasing number of quasars at high redshift is not due to dust obscuration \cite{Wall05}, 
the UV/optical spectral energy distribution (SED) remains a powerful diagnostic tool to advance the understanding of luminous QSOs. 
Thanks to the photometric data from SDSS, UKIDSS and WISE, the SEDs of the two BALQSOs can be plotted in the rest-frame after K-correction 
and Galactic extinction. We note that the radio flux at 1.4GHz of the two BALQSOs are below 1 mJy limit of the FIRST survey, in common sense, 
which means they belong to radio-quiet quasars and the bolometric luminosity should dominate their observational phenomena. Based on the 
optimized shifts using the average radio-quiet quasar template (the solid blue line, \cite{Elvis94}) for each source between the UV/optical 
and infrared bands, the most likely positions were plotted by differently dashed color lines as shown in Figure 3. 

Apart from apparent dropouts caused by the strong Ly$\alpha$ forest absorption in the higher redshift BALQSOs, the spectral profiles between 
1200\AA~ and 1600\AA~ of the two BALs are also deviated much to the up-shifted composite template, in which the discrepancies may be largely 
due to their evident absorption features, including the Ly$\alpha$+N V, Si IV and C IV or potential continua absorptions. Furthermore, this 
observational evidence, in part, accounts for the missing of their Ly$\alpha$ emission lines mentioned above. 
\begin{figure}[H]
\centering
  \includegraphics[width=9.0cm, height=6.5cm, angle=0]{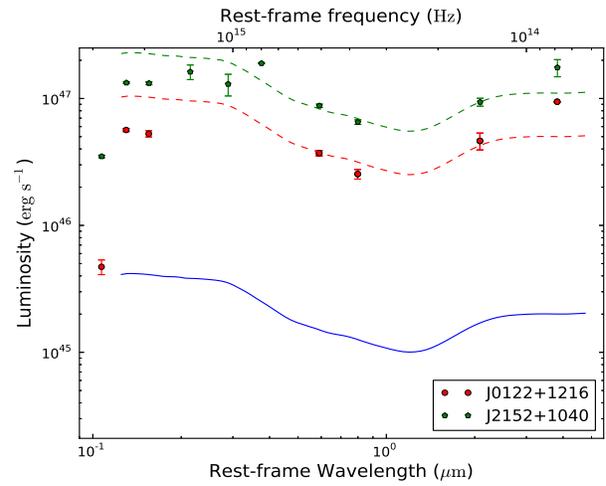}
  \caption{Rest-frame SEDs of the two BALQSOs at $z\approx 4.75$, which are compiled from multiple photometric surveys (2MASS, {\it WISE}\ and SDSS). 
  These data points are all calibrated in the rest-frame after K-correction and Galactic extinction correction. The blue solid line at 
  the bottom is the composite SED template of radio-quiet quasars in the local Universe \cite{Elvis94}, which is much less luminous than the two 
  high-redshift BALQSOs. Two dashed lines, shifted upward from the blue solid line, present the match of the template to the SEDs of two BALQSOs 
  between  1200\AA(10$^{15.398}$Hz)~ 
  and 4$\mu$m(10$^{13.875}$Hz)~ in the rest-frame wavelengths.}
  \label{Fig.1}
\end{figure}

\section{Summary}
Broad absorption line (BAL) outflows are observed as blueshifted wide troughs in the rest-frame spectrum of $\sim$20\% of 
quasars \cite{Hewett03,Ganguly08,Knigge08}. The energy, mass, and momentum carried by the outflows are thought to play a crucial role 
in forming the early universe and leading to its evolution (e.g., \cite{Scannapieco04,Levine05,Cattaneo09,Ciotti09,Ostriker10}). This 
has been further corroborated from the finding of a strong dependence between redshift and the BALQSO fraction, especially for the intrinsic 
C IV BAL phenomenon \cite{Allen11}. Based on the two BALQSOs discovered by us, their observational implications can be concluded as follows: 

1)The detached velocities of the C IV and SiIV lines are close to or might be beyond 0.1c, in combination with their very deep and broad absorption 
troughs, which strongly suggest spectacular outflow activities being violent during their evolution in the early Universe. The huge amount of outflows, 
in turn, should exert significant impact on the formation of BH and host galaxy.
 
2)All of their Ly$\alpha$ emission line peaks are obliterated to large extent in the spectra, which could be associated with the strong H I or N V 
absorptions around the inner circumnuclear region.

3)The extremely fast outflows among the two luminous BALQSOs, which are potentially driven by the super/sub-Eddington radiation, may be fundamental 
to or even dominate the process of their feedbacks. A better understanding of the kinematics through the combination of multi-wavelength data could 
shed light on the physical basis of the sub-relativistic activities among AGNs.

The discovery of two BALQSOs, according to their observational properties, would be particularly valuable to do some follow-up observations, 
especially for the near-IR spectroscopy, which can be fully used to investigate their true continua shapes, chemical abundances and AGN feedbacks, 
or more fundamental physics including the accretion disk and formation of black holes.

\vspace*{2mm} \Acknowledgements{\bahao We acknowledge the support of the staff of the LJT. Funding for LJT has been provided by CAS and the People's 
Government of Yunnan Province. X.-B. Wu thanks the support from NSFC grant 11373008. The work of J. M. Bai is supported by NSFC grants 11133006 and 
11361140347. This work is supported by the Strategic Priority Research Program \lq\lq The emergence of Cosmological 
Structures\rq\rq of the Chinese Academy of Sciences (grant No. XDB09000000).}

\end{multicols}

\end{document}